\documentclass[acmsmall,screen,nonacm]{acmart}
\usepackage{listings}

\setcopyright{none}
\renewcommand\footnotetextcopyrightpermission[1]{}

\begin{document}

\title{Beyond ChatBots: \name{} for Structured Thoughts and Personalized Model Responses}

\newcommand{\name}{\textsc{ExploreLLM}}

\author{Xiao Ma} 
\authornote{Author list is by descending order of contribution.}
\email{xmaa@google.com}
\affiliation{
  \institution{Google}
  \country{USA}
}

\author{Swaroop Mishra}
\authornote{Developed initial idea of prompt decomposition and tree-based design and interaction.}
\email{swaroopmishra@google.com}
\affiliation{
  \institution{Google Deepmind}
  \country{USA}
}

\author{Ariel Liu}
\email{arielliu@google.com}
\affiliation{
  \institution{Google}
  \country{USA}
}

\author{Sophie Su}
\email{sophiesu@google.com}
\affiliation{
  \institution{Google}
  \country{USA}
}

\author{Jilin Chen}
\email{jilinc@google.com}
\affiliation{
  \institution{Google}
  \country{USA}
}

\author{Chinmay Kulkarni}
\email{chinmay.kulkarni@emory.edu}
\affiliation{
  \institution{Emory University}
  \country{USA}
}

\author{Heng-Tze Cheng}
\email{hengtze@google.com}
\affiliation{
  \institution{Google Deepmind}
  \country{USA}
}

\author{Quoc Le}
\email{qvl@google.com}
\affiliation{
  \institution{Google Deepmind}
  \country{USA}
}

\author{Ed Chi}
\email{edchi@google.com}
\affiliation{
  \institution{Google Deepmind}
  \country{USA}
}

\renewcommand{\shortauthors}{Ma et al.}

\begin{abstract}
Large language model (LLM) powered chatbots are primarily text-based today, and impose a large interactional cognitive load, especially for exploratory or sensemaking tasks such as planning a trip or learning about a new city. Because the interaction is textual, users have little scaffolding in the way of structure, informational ``scent'', or ability to specify high-level preferences or goals. We introduce \name{} that allows users to structure thoughts, help explore different options, navigate through the choices and recommendations, and to more easily steer models to generate more personalized responses.
We conduct a user study and show that users find it helpful to use \name{} for exploratory or planning tasks, because it provides a useful schema-like structure to the task, and guides users in planning. The  study also suggests that users can more easily personalize responses with  high-level preferences with  \name{}. Together, \name{} points to a future where users interact with LLMs beyond the form of chatbots, and instead designed to support complex user tasks with a tighter integration between natural language and graphical user interfaces.

\end{abstract}



\vspace{-10pt}
  \begin{teaserfigure}
    \includegraphics[width=\textwidth]{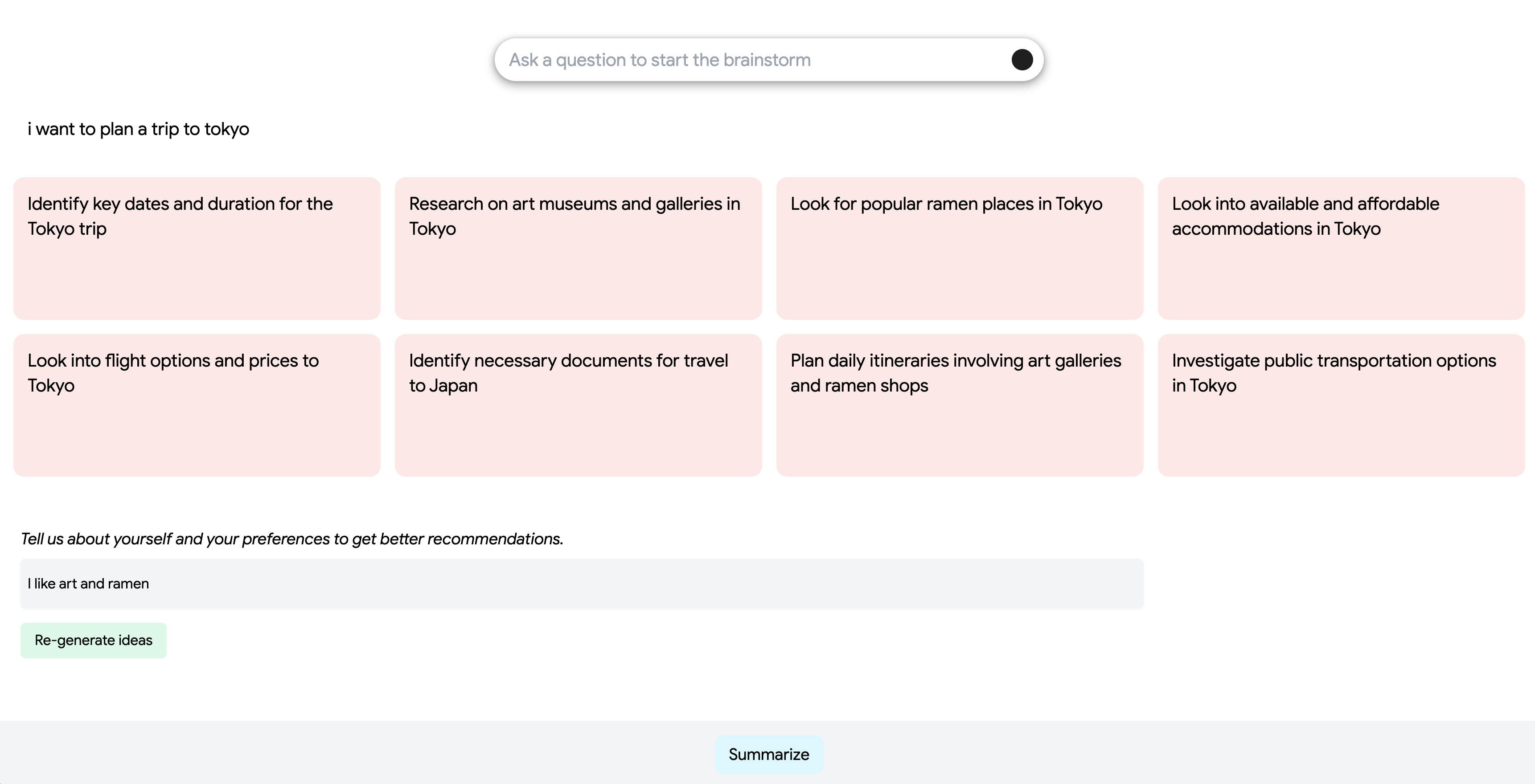}
    \caption{\name{} introduces a new interaction pattern with large language models (LLMs) by automatically decomposing complex tasks into sub-tasks, and allowing users greater task control and personalization.}
  \end{teaserfigure}

\maketitle

\section{Introduction}

Large language model (LLMs) powered chatbots have dramatically improved the user adoption of AI systems but have limited interaction patterns that are linear ands text-heavy.
Users can only carry out a single-stream conversation with existing chatbots such as Google Bard or OpenAI ChatGPT.
These chatbots, for the large part, respond with text\footnote{As of Nov 2023, the output of chatbots are becoming increasingly multimedia, but single-stream and text-heavy nonetheless.}. Such textual responses, which are often verbose, impose a significant cognitive load on users to understand and act upon, especially for complex tasks. Often, users have to engage in multi-turn conversations, both to probe what the chatbots can understand, and to communicate their intent and preferences. Similarly, because users can only respond with text, conversational repair~\cite{albert2018repair} is effortful. 

While there have been significant advances in prompt-based methods that unlock the reasoning and planning abilities of LLMs
~\cite{mishra2022reframing,wei2022chain,zhou2022least,yao2023tree,wang2023selfconsistency}, the interaction pattern between users and LLM-based assistants has largely remained the same.
This remains the case despite increasing evidence that users struggle to communicate complex tasks to assistants~\cite{kim2023understandingchatgpt,chang2023prompt}, much like conversational assistants that preceded them~\cite{zubatiy2023dontknowhelp}. 
Just as non-AI-experts use ad-hoc repair strategies to improve prompts for LLMs~\cite{zamfirescu2023johnny}, non-expert users similarly use ad-hoc tactics like adding details to their request, pointing out assistant errors in how the request was interpreted, or simply giving up on their original task and deviating to a related, simpler task~\cite{kim2023understandingchatgpt}.
These tactics sometimes act as a ``band-aid'', but still leave a majority of users unsatisfied~\cite{kim2023understandingchatgpt}.  More importantly, they prevent users from fully leveraging the potential of AI assistance to complete their tasks. 

In this work, we introduce a new interaction pattern between users and LLM-powered assistants, by combining a prompt-based \textit{task decomposition} method with a new schema-like graphical user interface (UI).
The new system, \name{}, decomposes tasks into sub-tasks \textit{automatically} using a prompt-based decomposition method.
Building on the LLM reasoning literature~\cite{wei2022chain,zhou2022least,wang2023selfconsistency,yao2023tree}, we custom-designed prompts to generate related and easier-to-solve sub-tasks related to the original user query.
Then, inspired by theories of schema in cognitive science~\cite{marshall1995schemas} and distributed sensemaking~\cite{fisher2012distributedsensemaking} in human-computer interaction, we render the generated sub-tasks for the users in a structured and interactive UI.
As a concrete example, for a complex task such as ``I want to plan a trip to Tokyo", \name{} will organize the query into
sub-tasks such as deciding on dates and duration, making hotel and flight arrangements, checking travel documents, etc. Such organizational structures, or schema, allow people to not only learn what aspects are important in the given task, but also act as a cue to express their own preferences. 
Further, we design a dedicated user preference solicitation UI and a recommender-like user interaction within sub-tasks, aiming at improving personalization and the ability of users to easily steer model responses.

We conducted a user study with eight participants where we asked users to compare ChatGPT and \name{} on a planning task. Our user study results show that \name{} is helpful in providing structured task breakdown which helps users to think and structure their thoughts. Users find the structured guidance that \name{} provides useful for planning. Users mention that \name{} is easier to personalize with their own preferences, in contrast to text-based chatbots.

The rest of the manuscript is organized as follows.
We first give an overview of the \name{} design, providing details on the key components and implementation details.
Through a qualitative user evaluation, we describe how it helps users complete complex tasks. In later discussion, we outline future work on integrating tool use and further opportunities for automation. 
Finally, we discuss the limitations of \name{}, some of which stem from foundational limitations of LLMs (such as hallucination), and others from the limited functionality that we developed.
We plan to open-source \name{} following the publication of this manuscript.

\section{Background}
\name{} builds on recent work in LLM reasoning, theories from human cognition, and prior work on natural language and graphical user interfaces in human-computer interaction (HCI).

\subsection{Prompting elicits reasoning and planning in LLMs}

In-context learning~\cite{brown2020language} and its evolution via various prompting methods have unlocked the reasoning and planning abilities of LLMs.
Once instruction-tuned~\cite{mishra2022cross, wei2021finetuned, ouyang2022training}, LLMs can follow specific instructions from end-users in natural language.
Leveraging the instruction following abilities of LLMs, researchers show that carefully designed prompts can improve LLM performance across a variety of reasoning and planning tasks, through intermediate representation in the form of thoughts~\cite{wei2022chain, nye2022show}, decomposition~\cite{patel2022question, khot2022decomposed,zhou2022least, press2022measuring}, search-decomposition mix~\cite{yao2023tree, saha2023branch}, structure~\cite{chen2022program, mishra2022lila, gao2023pal}, abstraction~\cite{zheng2023take} and optimization~\cite{yang2023large}. 
Prompting based solutions have significantly improved model capabilities across a diverse range of tasks such as program synthesis~\cite{kuznia2022less}, dialogue~\cite{gupta2022instructdial}, biomedical applications~\cite{parmar2022boxbart}, style transfer~\cite{reif2022recipe}, multilingual~\cite{shi2022language}, moral reasoning~\cite{ma2023let} and a large spectrum of natural language understanding tasks ~\cite{wang2022self, madaan2023self}. 
However, most of these advancements are targeted at improving LLMs' performance on some benchmark tasks, rather than benefiting end users (non-expert users in particular~\cite{zamfirescu2023johnny}).
In this work, we explore the possibility of leveraging prompt-based methods to better support end users in their complex exploratory or planning tasks.

\subsection{Schemata support thinking and problem solving}
Interestingly, some methods for eliciting the reasoning ability in LLMs have roots in psychology and cognitive science -- particularly the concept of \textit{schema}.
A schema is a framework, outline, or plan for solving a problem~\cite{marshall1995schemas}.
Prior work shows that schema is an effective tool in supporting human problem solving~\cite{powell2011solving,ma2020domaindiagrams}.
One way to create schemata is through task decomposition.
The intuition of breaking tasks down also aligns well with the distributed sensemaking work, which indicates that in solving complex problems, it is useful for people to have a starting problem structure, that they can later customize to their own goals~\cite{fisher2012distributedsensemaking}.
Further, although originally developed by cognitive scientists to describe human learning~\cite{thorndyke1979schema}, the concept of schema has inspired subsequent frameworks of machine intelligence, such as work by Minsky on \textit{frame} and \textit{frame-systems}~\cite{minsky1974framework}.
When faced with a new situation, humans select from memory a structure called a \textit{frame}, a data-structure representing a stereotyped situation.
Collections of related frames are linked together into \textit{frame-systems} that can be manipulated while allowing information coordination~\cite{minsky1974framework}.
In this work, we leverage LLMs' reasoning ability to assist humans in creating schemata for problem solving using prompt-based task decomposition automatically.
In addition, we draw inspirations from frames and frame-systems when designing system interfaces.
Users can interact with each sub-task as connected components via a UI for further customization and the \name{} system keeps track of user contexts and preferences for coordinated decision making.

\subsection{Why natural language alone is not enough}

Natural language user interfaces (NLUIs) and graphical user interfaces (GUIs) are two major ways for humans to interact with machines.
As early as 1972, Winograd developed SHRDLU, a natural-language interface that manipulates blocks in a virtual ``blocks world''~\cite{winograd1971procedures}. The effectiveness of NLUIs are limited by the capability of underlying AI systems.
The invention of GUIs in the 1970s was largely a response to the lack of the natural language understanding and generation abilities of machines.
GUIs played a major role in the wide adoption of personal computing~\cite{jansen1998graphical}.
With the recent advancements in LLMs, NLUIs received renewed attention in the form of chatbots.

At the same time, there is compelling evidence that natural language interfaces alone are not enough: decades of work in cognitive science suggests that thinking is intimately tied to \textit{doing}, not just speaking. Put differently, thinking is a process that is not limited to what happens in our brain, but instead it is ``distributed'' throughout our environment~\cite{hollan2000distributedcognition}. For example, people find it much easier to move lettered tiles into various arrangements for playing Scrabble~\cite{maglio2020interactivescrabble} 
(rather than merely talk through alternatives), and professionals frequently draw rough diagrams to ``help them think''~\cite{ma2020domaindiagrams}.

While natural language is flexible, it limits users mostly to single-stream and text-heavy interaction patterns.
GUIs have unique advantages that are more compatible with human cognition and sensemaking.
Consistent with the schemata concept, graphical user interfaces are particularly helpful in adding structure to a task, and allowing users to notice aspects of the task that matter the most~\cite{suwatversy2001seeingsketches, gegenfurtner2019learningvision}. For instance, a table comparing alternatives across the most important dimensions  helps programmers choose between competing technical approaches~\cite{liu2019unakiteSOtable}, and a ``mind map'' developed by other users can help users learn important aspects of a complex problem~\cite{fisher2012distributedsensemaking}.
Over time, if interfaces are well-structured and  predictable, users may develop tool-specific expertise in using them that extends beyond their conceptual task understanding~\cite{klemmer2006bodies}.
For example, photo-editing experts develop expertise with specific tools like Adobe Photoshop that goes beyond a conceptual understanding of editing photos.
Unfortunately, current AI chatbot interfaces do not take advantage of GUIs, and instead generate responses to each query linearly.
As a result, users struggle to develop a strong mental model of such interactions, which are especially important for complex tasks.

In this work, we introduce \name{} to automatically induce a structure that highlights salient dimensions for exploratory tasks, by combining the best of natural language and graphical user interfaces.
Some recent work has started exploring the better design of graphical interfaces for LLMs~\cite{suh2023sensecape,jiang2023graphologue}.
For example, Sensecape is an interactive system built with LLMs to support flexible information foraging and sensemaking~\cite{suh2023sensecape}.
Graphologue converts the text-only response from LLMs to interactive diagrams to facilitate information-seeking and question-answering tasks~\cite{jiang2023graphologue}.
\name{} builds on these initial explorations to create predictable, structured interfaces for complex exploratory tasks, focusing on task decomposition.
Together, we demonstrate a promising direction of ``hybrid'' user interfaces, where a tighter integration is drawn between natural language and graphical user interfaces: language allows for expression of complex and abstract goals and preferences, and graphical representations allow for more structured understanding and exploration~\cite{kirsh1995coincounting,klemmer2006bodies}.

\section{Methods}
In this section, we first outline the key design components of the \name{} system and implementation details.
Then we describe the setup of the user study.

\subsection{\name{} Overall Design}
We first provide an overview of the system, and then delve into details and the design rationales of each component. In this work, we design a standalone prototype for simplicity. However, this mode of interaction can potentially be invoked from the current chatbot interfaces with magic commands such as ``/explore''.

\subsubsection{System Overview}

\begin{enumerate}
    \item \textbf{Node:}
    The \name{} system has an underlying tree-like data structure.
    Unlike traditional chatbots, we create an abstraction of a node that can be nested.
    A node is a unit of interaction and can represent different forms of interactions (e.g., multi-turn natural language chats or UI interfaces).
    By default, a new node is created when the user starts interacting with the system.
    When needed, the system automatically creates children nodes for users to explore through task decomposition.
    All nodes form a tree-like hierarchical structure that holds all the context of the user's exploration journey. For this work, we limit the nodes to depth = 2 for simplicity (root user query with one layer of children nodes). In the future this can be extended to more layers or graphs.
    
    \item \textbf{Personal Preferences:} At any given node, the users can provide free-form context that the system should be aware of for better personalization. The personal preference context is shared globally across all nodes.
        
    \item \textbf{Options:} In each node, the system will take personal preferences into consideration and dispatch a backend call to present some options for the user to choose from.
    Users can interact with different options via a checkbox UI to indicate their preferences.
    
    \item \textbf{Summarize:} After sufficient exploration, users may want to tie everything back together and get a summary of their journey so far.
    Therefore, the system has a ``summarize'' function that is available on each page.
    Users can click on the button to exit to the root node and get a text summary of their entire interaction across the system.
\end{enumerate}

\subsubsection{Node}
One of the most important design goals of the \name{} system is to better support complex and under-specified tasks that require exploration.
We address the challenge that complex tasks require high cognitive load by creating a tree-like abstraction.
Through reasoning literature, especially on decompositions~\cite{zhou2022least, patel2022question, khot2022decomposed}, we know that LLMs are capable of decomposing a complex problem into a list of easier subproblems.
In this work, we leverage task decomposition abilities of LLMs for \textit{users' benefit}, rather than as a method to improve LLM's task accuracy.
Users can type into a generic query box and the system will first create a root node for this query, and then automatically calls the LLM task decomposition endpoint to create a list of easier sub-tasks.

We use the prompt in Figure~\ref{fig:prompt-subproblems} for automatic task decomposition.
\begin{figure*}[ht!]
\begin{lstlisting}[breaklines=true, numbers=none, identifierstyle=\textnormal, basicstyle=\tiny\ttfamily, frame=single]
I want to accomplish the main goal of: {text}
To better assist me, please break down the problem into sub-problems.
Each sub-problem should help me to solve the original problem.
Make it so that each sub-problem is not trivial and can be helpful.
Take my context and personalization cues to personalize the sub-problems.
Make sure each sub-problem is concise and less than 15 words.

Personalization Cue: {selected_options}
My Context: {user_context}

Output format (make sure only output a valid JSON object that can be parsed with javascript function JSON.parse).
Do not include any '``` or json'.
{{
  "sub_problems": A list of strings (max 8), each as a valid sub-query
}}

Output:
\end{lstlisting}
    \caption{Prompt used in \name{} to break a complex task down into structured sub-tasks.}
    \label{fig:prompt-subproblems}
\end{figure*}
\begin{figure*}[ht!]
    \centering
      \frame{\includegraphics[width=\linewidth]{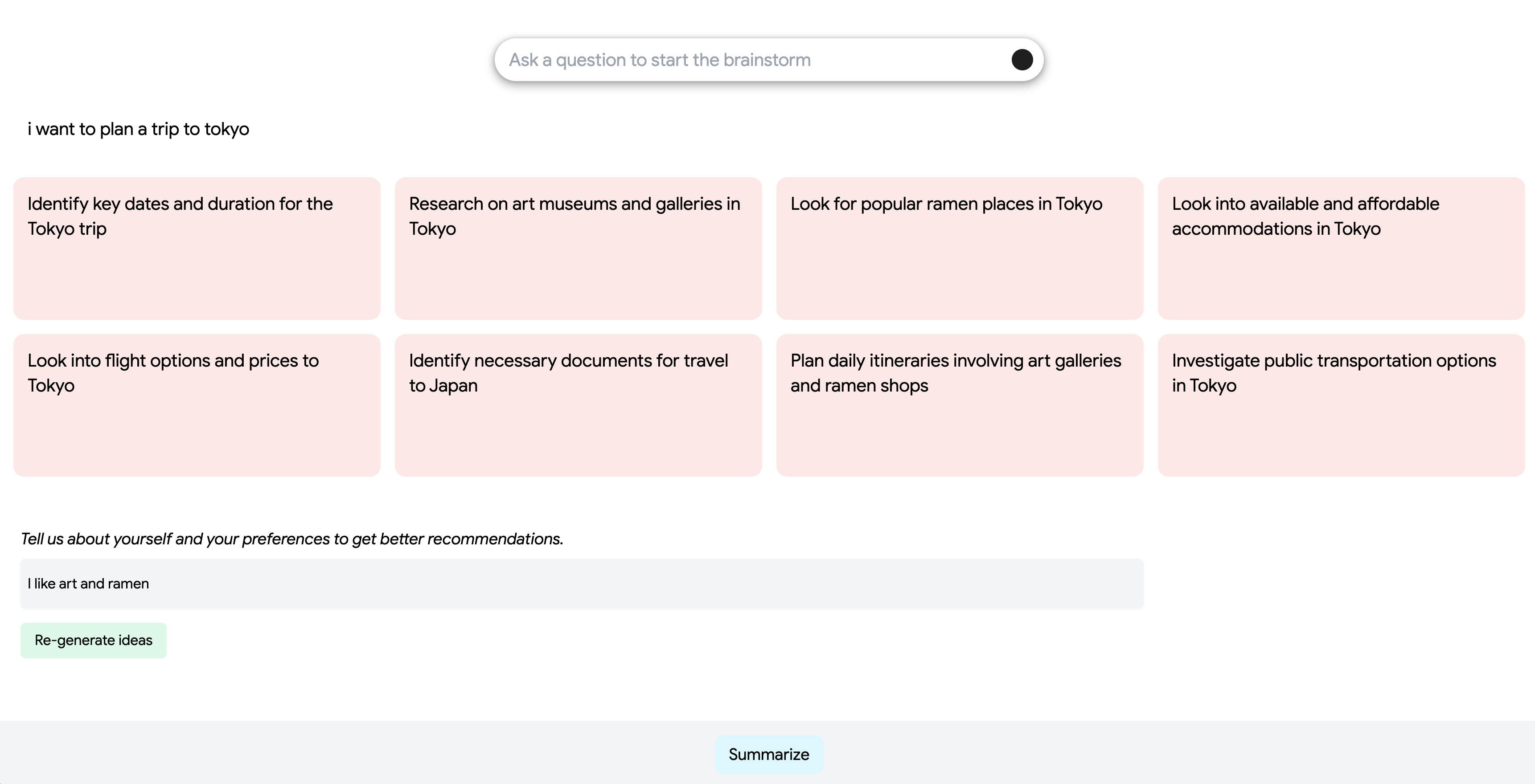}}
    \caption{
    User starts interacting with the \name{} system by typing a query. The system automatically breaks down the original user query into sub-tasks using a custom prompt, and then create nodes that represent each sub-task. Users can see the sub-tasks rendered as ``cards'', and interact with each one. A dedicated UI prompts the users to specify personal contexts and preferences for personalization. Once user is satisfied with the exploration, they can click the summarize button at the bottom and \name{} generates a summary of the user journey.
    }
    \label{fig:system}
\end{figure*}
The original user query and any additional personalization cues are passed to the LLM through this prompt (Figure ~\ref{fig:prompt-subproblems}).
The output of the prompt is a list of sub-tasks that, together with the original task, forms a hierarchical tree-like structure.
We display each sub-task to the users as ``cards'' that they can interact with (see Figure~\ref{fig:system}), mimicking a schema-like structure.
When users hover over one of the card, a ``see more'' button appears.
Like a discussion thread in online forums, users now can focus on each task separately while the system keeps track of the logical structure, thus reducing user's mental load.

\subsubsection{Personal Preferences}
The second design goal of the system is to support better personalization.
Complex tasks usually have important personal contexts or constraints that are important to the user, and are incredibly overwhelming to elicit in one go.
LLMs are trained to be ``general purpose'', which dictates that the responses are often tailored to an ``average user''.
This ``regression toward the mean'' causes the LLMs response to be generic and not personalized.
While it is possible for LLMs to offer more personalized responses once the users clearly specifies their preferences, prior work in recommender systems show that users themselves often are unaware of their needs and often struggle to express them, especially in the beginning of a planning process~\cite{neidhardt2014eliciting}.

To more prominently elicit user preferences, we include a dedicated UI that is always available for users to update their preferences.
We prompt the user to ``tell us about yourself and your preferences to get better recommendations''.
Importantly, the personalized context field is always available regardless which node the user is exploring, so that they can update the context if the task at hand reminds them of some preferences.
The personalized context is passed in the prompt to the LLMs in all subsequent backend calls for better personalization.

\subsubsection{Options}
Another design solution for reducing mental load while increasing personalization is providing users options to choose from.
Again, we draw inspirations from recommender systems.
Prior work notes that the cognitive load for users to provide accurate preferences and ratings for items is much greater than providing implicit feedback (e.g., selecting an option they prefer)~\cite{oard1998implicit}.
We leverage this insight and construct a LLM options generation endpoint with the prompt in Figure~\ref{fig:options_prompt}.

\begin{figure*}[ht!]
\begin{lstlisting}[breaklines=true, numbers=none, identifierstyle=\textnormal, basicstyle=\tiny\ttfamily, frame=single]
User: {text}

== Instructions ==
The user wants to: {context}
Here is one of the sub-query to help answer the main query.
Go into details to help me with the sub-query.
Show me some options to personalize and choose from.
Be concrete and make sure the options are valid choices to finish the task in sub-query.

Personalization Cue: {selected_options}
My Context: {user_context}
When coming up with options, make sure they are diverse and representative of multiple demographics, cultures, and view points.

Output format (make sure only output a valid JSON object):
Do not include any '``` or json'.
{{
    "recommended": Your recommendation. ,
    "options": A list of options (at least 5) for me to choose from. Each option is a single string. Provide helpful details. Don't include numbers or bullet points.
}}

== End of Instruction ==

User: {text}
Output:
\end{lstlisting}
    \caption{Prompt used in \name{} to provide options for users to choose from.}
    \label{fig:options_prompt}
\end{figure*}

We included the request for ``diverse and representative'' inspirationally to make sure the options generated are inclusive~\cite{lahoti2023improving}.
However, due to the small sample size of our user studies, we did not formally evaluate whether such a prompt is effective in increasing the options diversity or swaying user preferences.
We discuss this in more detail in limitations and future work.

\begin{figure}[h]
    \centering
      \includegraphics[width=\linewidth]{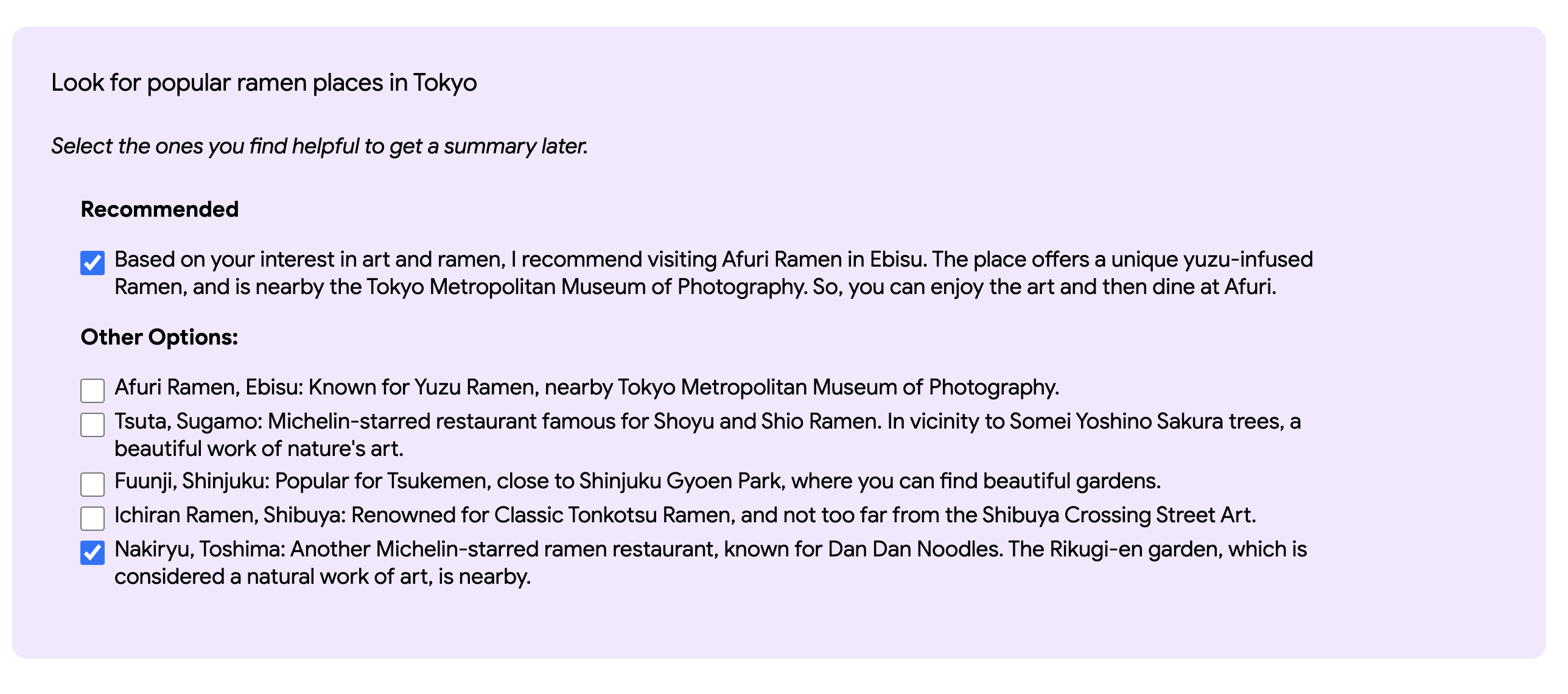}
    \caption{\name{} generates personalized options for users to choose from.}
    \label{fig:options}
\end{figure}

Once the users click on a node to ``see more'', the system redirects them to a whole page screen to focus on the sub-task at hand.
In the backend, the system dispatches the call to the LLM options generation endpoint. One the results come back, we display the generated options with a checkbox UI (see Figure~\ref{fig:options}).
The checkbox UI is designed to make it as easy as possible for users to provide implicit feedback through selection for better recommendations.
The system keeps track of user selections, and any subsequent prompts to the backend will include all user selections for better personalization.
The design intends to resemble a more passive ``browsing'' experience to minimize the mental load for users.

\subsubsection{Summarize}
\begin{figure*}[ht!]
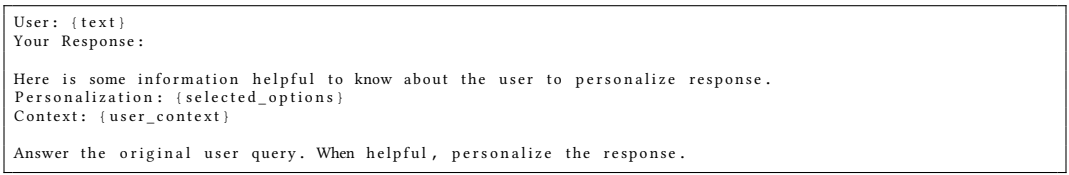

\begin{lstlisting}[breaklines=true, numbers=none, identifierstyle=\textnormal, basicstyle=\ttfamily\tiny, frame=single]
User: {text}
Your Response: 

Here is some information helpful to know about the user to personalize response.
Personalization: {selected_options}
Context: {user_context}

Answer the original user query. When helpful, personalize the response.
\end{lstlisting}
    \caption{Prompt used in \name{} to generate personalized summary across different sub-tasks.}
    \label{fig:summarize_prompt}
\end{figure*}

After sufficient exploration in each sub-task, users may want to come back to the main task and get a summary of their journey.
Therefore, we include an UI for summarization.
Once the user clicks on ``summarize'', the system gathers all user interaction signals and passes it to the LLM summarization endpoint with the prompt in Figure~\ref{fig:summarize_prompt}.
The system redirects the user back to the root node and produces a textual summary of the main task.

\subsection{Implementation}
We implemented the above design with Next.js 13 for the front-end, and OpenAI GPT-4 API and FastAPI as backend.
It is important to note that the interaction design is agnostic to the LLM API of choice, as long as the LLM has sufficient instruction following and reasoning capabilities.
User data is stored in a secured MongoDB database and only selected authors have access to the database.
The services are hosted on Google Cloud Run and Vercel.
We plan to open-source the implementation of \name{} following the publication of this manuscript.

\subsection{User Studies}

We conducted a qualitative study to evaluate \name{}.
We used convenience sampling for recruiting participants.
There was no restriction on age, gender, and the language was limited to English.
We asked a few screening questions about prior experiences and attitudes towards AI chatbots, such as Bard by Google, ChatGPT by OpenAI and Claude by Anthropic.

Then, participants were instructed to do a two-part unmoderated user study.
We asked the participants to record themselves during the study, while presenting their computer screens and thinking aloud.
Participants were informed of their rights and signed a standard consent form before the study.
We issued a small token of appreciation as incentive for the study without monetary payments.
For two of the first studies (P1 and P6), one of the authors of the paper was on the same call as an observer with the participant, and provided light clarification assistance when needed.
We made minor changes to the instructions to ensure subsequent unmoderated studies can proceed smoothly.
All studies were conducted during Nov 6 - 10, 2023.

The study was within-subject and counterbalanced.
For each part of the study, participants were instructed to do a travel planning task to a destination of their choice, one using ChatGPT, the other using \name{}.
We always instruct the participants to plan the trip to the same destination in the second task as the previous one.
We provided a login with ChatGPT plus subscription and participants were instructed to use ChatGPT-4\footnote{Note during the study period the ChatGPT system was changed from ChatGPT-4 to turbo. We do not know the details of the change but in ChatGPT-4 turbo, the system also has web browsing capability, which resulted in some of the participants commenting about transparency of the system.}.
We randomized the order of the system the participants use.
In the end, we had five participants using ChatGPT first (P1-5), and three participants using \name{} first (P6-8).

We collected age and gender as demographic information in the exit survey.
We transcribed the user study videos and two authors of the paper conducted qualitative coding on the transcripts to identify common themes, first independently, and then discussed and agreed on the final findings.

\subsubsection{Main Task}
We provided the following instructions for the main task.
It is important to note that our system design is not limited to travel planning.
We chose to evaluate trip planning as an example task because it is general, typically complex, and allows for individual user constraints and preferences.

Key study instructions are as follows:

\textbf{Imagine you are planning a trip to a destination of your choice using [system name].}
\begin{itemize}
    \item You have less than~5 min to complete this task.
    \item You can have as many interactions as you like using this system.
    \item Please make sure this trip is personalized to you (e.g., cost, time, location).
    \item Don't forget to think aloud as you try to complete the task!

\end{itemize}

\textbf{Once you have completed the task, answer these following questions verbally:}
\begin{itemize}
    \item Tell us in detail, what do you find most helpful and unhelpful from this result? 
    \item If at all, which part of the result do you find personalized to you? 
    \item If at all, how much does this system make you feel more or less confident about planning a trip? 
    \item Is there anything that you would like to comment about this task?
\end{itemize}

\subsubsection{Participant Demographics}
Eight participants took part in the study, out of which six identified as male, one female, and one preferred not to say.
Seven out of eight participants are within the 25-34 years old, with one participant 35-44 years old.
Seven participants are located in the U.S., with one in the U.K..
We did not collect other demographic information for privacy due to the limited sample size.
We acknowledge that the lack of diversity of participants is a significant limitation, and discuss future work to address this problem in later sections.

\section{Results}

Overall, participants confirmed our hypotheses that the current chatbot system provides generic and verbose responses, and that they liked \name{}'s ability to provide structured task breakdown and personalization, despite some usability issues.
Importantly, we also found that hallucination is a major hurdle in building trust in the system.
Participants often pointed out where the information provided in the system is wrong, or that they don't trust the information and need to conduct their own research for additional verification.
Finally, participants expressed wishes for more control of the system, richer content and tool use, which we discuss in future work.

Below we report in more detail each of the findings.
We refer to a particular participant as P\#.

\subsection{Limitations of the Chat-Only UI}

Participants pointed out two major limitations of the current chat-only UI for exploratory tasks.
All but one participant mentioned that ChatGPT responses are  verbose and generic.
Multiple participants used the ``stop generating'' button to interrupt ChatGPT mid-response (while the response is streaming), commenting \textit{``this is really verbose''} (P5).
P3 stated, \textit{``I didn't need this much text to start with.''} P7 said, \textit{``too much, stop it''}.
Often, people interrupt the generation when they notice the response being streamed back is off track.
After the interruption, people often ask follow up questions to steer the conversation to another direction.
For example, P5 asked \textit{``Can you tell me when is the best time to go to [destination]?''} after noticing the ChatGPT was going on and on about the cost of the trip after the initial prompt.

The second major limitation of the chat-only interface is that responses are generic.
P1, P2, P5 and P6 commented that ChatGPT gives generic answers.
P1 stated, \textit{``It was giving me kind of generic stuff. I had to remember to tell it stuff about me.''}
P6 mentioned that \textit{``It was not really personalized because this is just giving me information.''}
Even when the user provides more personal preferences, sometimes ChatGPT fails to take it into consideration when generating responses.
For example, P5 specifically mentioned something that they are not interested in but the system still generates responses containing those options, revealing a lack of steerability.

\subsection{Structured UI and Guided Task Flow}
Compared to the chat-only UI of ChatGPT, participants liked the structured UI of \name{} and the guided task flow.

In particular, participants found that the broken down list of sub-tasks was useful and helped them think, plan and navigate.
P5 stated, 
\textit{``Having these different aspects of the trip surface upfront is helpful [...] where was the best time, how do we get there, what are some things to do there? It's great that the system surfaces these up front. So I don't have to like consider all the different aspects because I feel like I will miss things.''}
P2 stated, \textit{``the logical flow was very clear.''}
P3 said, \textit{``the whole nature of guiding people through their problem is a big one.''}

People also liked the fact that the list of sub-tasks persists as a collection of artifacts to reference back on.
Compared to the chat-only interface, P1 stated, \textit{``It's just nice not have a wall of text to have to look back through.''}
In contrast to the wall of text of ChatGPT, people found \name{}'s UI glanceable and that it keeps a better record of the task state.
P2 commented on the list of sub-tasks, saying that \textit{``I really like the list [...] I can see those lists so I felt more confident.''}
This is in contrast to the linear chat-only UI, where the users end up with a long chat history and have to \textit{``dig back through to find the interesting parts''} (P1).

At the same time, users indicated their wish for more control.
P1 indicated that \textit{``I wish that I could [...] X out some of these things. [...] these seem pretty similar so I could like close one of those so I can remember [..] like keep the high priority ones open.''}

\subsection{Personalization}

Contrary to the generic responses in chat-only interactions, the personalization aspect of \name{} stood out to participants.

Participants found that it was easier to get personal preferences into the system because of the dedicated UI (P1, P2, P3, P6).
P3 said, \textit{``I do think this is more personalized because it was easier to get that out of me.''}
P2 also noticed that the personalized context carries over globally throughout the entire interaction. \textit{``In the middle of the tasks, I specified clean and budget and it carried over to other tasks, and it helped me to narrow down those options and prioritize.''}

Nonetheless, participants expect \name{} to be more proactive in eliciting their personal preferences, by doing \textit{``more cognitive lift''}.
When commenting on the options generated by the \name{} system for flights, P3 added,
\textit{``It should have probably asked me [...] where do I have airline miles [...] or am I a member of a thing? Do I have any preferences? I hate having to remember what to tell it. If I talk to a travel agent, I don't get that. [...] They know what to ask to start. So that's a pretty big difference.''}

\subsection{Options as Responses}

Five out of eight participants (P1, P2, P5, P6, P8) indicated liking having a list of options to choose from in \name{}.
For some participants, the benefit mainly comes from reducing mental load.
P8 said, \textit{``I think the idea of specifying sort of vaguely what you're interested in, getting some options, picking one, and then having the other follow-up responses take those options into account make sense.''}
P2 added, \textit{``It removed all the burdens and it'll is just basically gave me options that I just need to choose from so it was super helpful.''}
For others, such as P6, the benefit comes more from the ability to personalize.
P6 saw different layover options for the flight in one of the sub-tasks, and pointed out that having different stopover options is important because sometimes there are visa limitations and they prefer to go through one of the options than other ones.
Finally, people also liked the explainability of the options, indicating that the explanations associated with each option are helpful.

However, issues arise when there are similar or too many options.
P5 observes that for the subtask ``Check the best time to visit [destination]'', \textit{``the recommended and other options are pretty much the same''}. (The options were all in autumn, including late August, September, and early October.)

Notably, many participants expressed the desire for richer content (P5, P7) and tool use (P1, P2, P6).
P5 suggested wanting to see the list of options on a map to visualize their relationships to each other and how close things are.
\textit{``Some richer content like images and Maps should help and if I could [...] click on this to [...] get more details about the champagne houses or click on the locations to understand more about [...] the scenery around it, that would be helpful.''}
P5 also switched tabs to search for some suggested destinations on Google maps mid-task.
P7 also suggested that they want to see the full itinerary on a map.

P2 suggested connecting to external tools would be more helpful.
When the system was loading, P6 speculated that the system was using some tools to search for flight options (the system was not using any tools yet and we discuss this in future work).
Participants also expressed the desire for the system to take actions rather than providing information alone.
P1 mentioned that even though the system provided information, \textit{``I still have to do the annoying nitty-grittys like actually book the tickets and figure out calendar dates''}.

\subsection{Hallucination}

Across both ChatGPT and \name{}, participants noted hallucination as a major limitation and expressed reservations in trusting the results fully.

Participants noted multiple times that some information generated by ChatGPT and \name{} is wrong.
P2 requested travel planning for the NeurIPS 2023 conference in \name{}.
Although the system correctly breaks down the task into first looking into the location of NeurIPS conference, it tried to look up the location for 2022 rather than the latest.
Further, the system hallucinates and thinks that the NeurIPS 2022 was in Vancouver, Canada. P2 was confused and looked up NeurIPS 2022's location, which is New Orleans, USA.
Factual errors like this get propagated to downstream tasks such as hotel booking and is very disruptive to the entire experience.
In another example, P6 noticed that one of the suggested airports for flight options did not exist.
The \name{} system suggested one option to go from London to India as ``Emirates EK6\footnote{Fact checking shows that Emirates EK6 leaves from London Heathrow to Dubai.} flight departs from London Gatwick and arrives at Kolkata Bhawanipur with one stop at Dubai''.
P6 commented that \textit{``This is wrong, there is no airport at Bhawanipur.''}

Such hallucination has significant safety implications and also dramatically limits the system's usefulness and the extent to which users trust the system to take actions autonomously.
P1 commented on one of the ChatGPT responses, \textit{``I don't know if I would trust it even it's telling me [...] be careful about the transit options being canceled because of snow, I think I need to do some research into exactly what that means''.}
We discuss this further in the limitations and future work section, including opportunities for better grounding, tool use, and integration with other systems.

\subsection{Usability Issues}
As a research prototype, some usability issues have emerged in \name{}.
Most notably, all participants noticed that the system has high latency.
Almost all participants noted that the system is slow when generating different options.
Because the response back from the prompts are structured json data, it is not very suitable for streaming, which could have reduced the perception of latency.
As the underlying LLMs become faster, the latency issue can be mitigated to a large extent.

Providing more transparency about the process to the users can be another design solution for handling latency.
When the system is loading, currently \name{} indicates ``thinking''.
P3 wondered, \textit{``What is it thinking about? ChatGPT with browsing is a little bit clearer because it's telling me what tools it's using and how it's thinking about stuff.''}
As more tools and functionalities get integrated into the system, revealing the inner workings of the system is important for user transparency and trust.

\section{Discussion}
In this work, we introduced the \name{} system designed to provide more structured task guidance and easier personalization.
Echoing findings in recent work~\cite{suh2023sensecape,jiang2023graphologue}, our user studies support the motivating hypothesis that current chatbots' responses can be verbose and generic.
Participants liked having the structure of sub-tasks and the ability to personalize, wishing for even more control.

One of the most important findings of our work is that much of the prompt engineering work and the ``thoughts'' of LLMs can have direct user benefits when appropriately exposed to the end users.
This intuitively makes sense as many underlying structures of LLM reasoning methods are compatible with how humans think and solve problems.
We show that prompt-based methods can effectively aid humans in creating schemata (or ``mental models'') for problem solving.
With the scaffold of a logical task structure, each sub-tasks can be explored separately to reduce cognitive load, while loosely coupled together for effective information coordination across the system.

In addition, our findings show that task decomposition is promising for better tool use.
LLM assistants do not exist in isolation, and users wish for a tighter integration with existing data and tools.
Tool use is especially important given that hallucination presented itself as a major hurdle in gaining user trust.
Through task decomposition, we can break down a much more complex task into concrete sub-tasks that have readily made tools suitable for the sub-task (e.g., checking the weather of a particular location; searching for a particular type of place on maps).

We see that participants intuitively begin to expect or wish for better tool use once they start exploring sub-tasks.
The tree-like nature of \name{} tasks system makes it highly compatible for deeper decomposition and integration with external tools.
For example, one of the sub-tasks generated by the \name{} system for travel planning is flight booking.
This sub-task can be further broken down into sub-sub-tasks, such as deciding on dates and specifying departure locations.
Once the user interacts with the sub-sub-tasks, we can map user input to parameters when invoking specialized tools such as flight search engines to provide accurate and personalized results.
For complex open-ended tasks that require extensive planning, a related open problem is how to decide when to use tools autonomously to solve sub-problems, and when to elicit user feedback.

More generally, our work shows the promise of re-imagining the relationship between natural language user interfaces (NLUIs) and graphical user interfaces (GUIs)~\cite{jansen1998graphical}.
As much as people are excited about the giant leap in NLUIs, there lies an opportunity to re-imagine the design of GUIs under this new interaction paradigm. \name{} offers one exploration of the possibilities in creating new ``hybrid'' systems that integrate the best parts of both natural language and graphical user interfaces.

\section{Limitations and Future Work}
One of the most significant limitations of our work is the lack of diversity in participants.
In addition to convenience sampling, our participants have limited representation in terms of age and gender, and are biased towards experienced users of AI chatbots.
Five of our participants reported interacting with AI chatbots a few times a week, and the rest three reported using it every day.
We plan to expand participant diversity in a follow up work.

Secondly, we only explored one layer of task decomposition without tool use or data integration.
Future work can extend to more layers of task decomposition and integrate existing tools to sub-tasks, or even explore leveraging the tool making abilities of LLMs itself~\cite{schick2023toolformer,cai2023large}.
Our system only elicits user context through input text, and future work can explore pulling user contexts directly from user data and external apps, such as emails and calendar (e.g., what reservations users already made, what events are on the calendar, what are the dates of the travel).
In addition, we only explored a checkbox UI and displayed options in sub-tasks as a list.
Future work can explore richer UI as participants mentioned, such as maps and diagrams.

The prompt we used for task decomposition and options generation endpoints can be further tuned for quality and diversity.
For example, we can tune both prompts to avoid repetition or options that are too similar.
It is also important to consider fairness in options generation given prior work on algorithm fairness~\cite{corbett2017algorithmic} and the impact on user behaviors by social media ranking algorithms~\cite{eslami2015always} and to guard against overreliance~\cite{openai2023gpt4}.
While we included the prompt to generate ``diverse and representative'' options, we did not formally evaluate the diversity of options generated due to the limitation of sample size.
Future work could formally evaluate the impact of prompts on the diversity and quality of options generated, and their downstream impacts on user behavior.

Finally, as a research prototype, our implementation still has usability issues.
We plan to open-source our implementation following the publication of the manuscript to allow the community to build and iterate on \name{}.

\section{Acknowledgement}
We thank Jim Maddock for designing the initial version of the user evaluation plan and literature review. We also thank Melvin Johnson and Varun Godbole for their insightful feedback during the early phase of this project.

\section{Author contribution statements}
Xiao led the overall project, co-designed the personalization, options generation, and summarization interactions, implemented the system, and conducted user studies and analysis.
Swaroop developed the original idea of tree-based interaction, came up with the initial design, demo video and write up.
Ariel contributed to user studies and interaction design.
Sophie iterated on the initial demo and designed the system UI and interaction.
Jilin contributed to the paper framing and user studies design.
Chinmay participated in early discussions about the LLM sub-task decomposition method, and helped with theoretically grounding work. 
Heng-Tze and Quoc provided continuous advice and feedback on the research of prompt decomposition and tree-based user-LLM interactions.
Ed provided feedback on paper framing and connection to prior literature.
All authors reviewed the manuscript.

\bibliographystyle{ACM-Reference-Format}
\bibliography{ref}

\newpage
\appendix

\section{Appendix}

\begin{figure}[h]
    \centering
      \includegraphics[width=\linewidth]{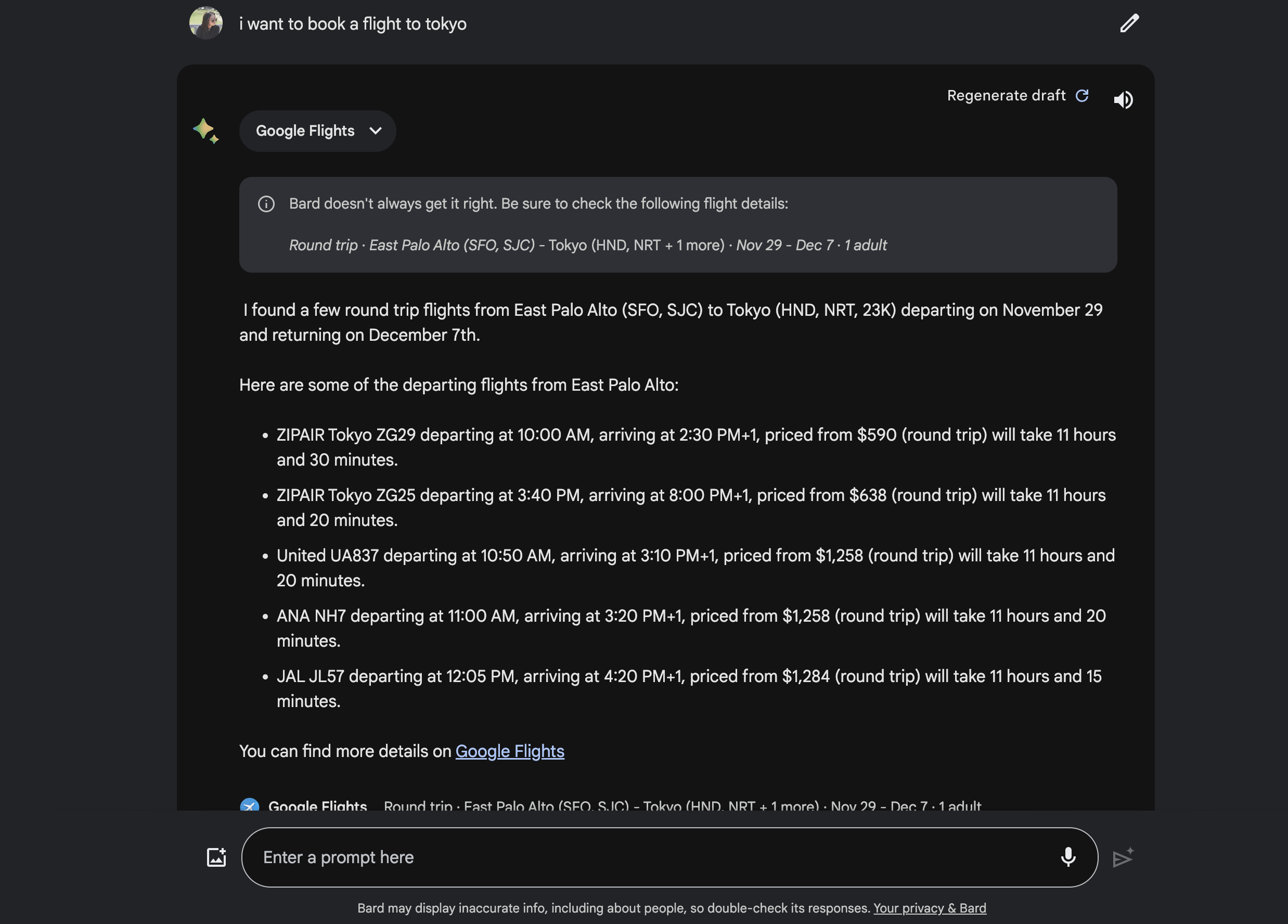}
    \caption{Bard results for the query ``I want to book a flight to Tokyo''}
    \label{fig:bard_flight}
\end{figure}

\begin{figure}[h]
    \centering
      \includegraphics[width=\linewidth]{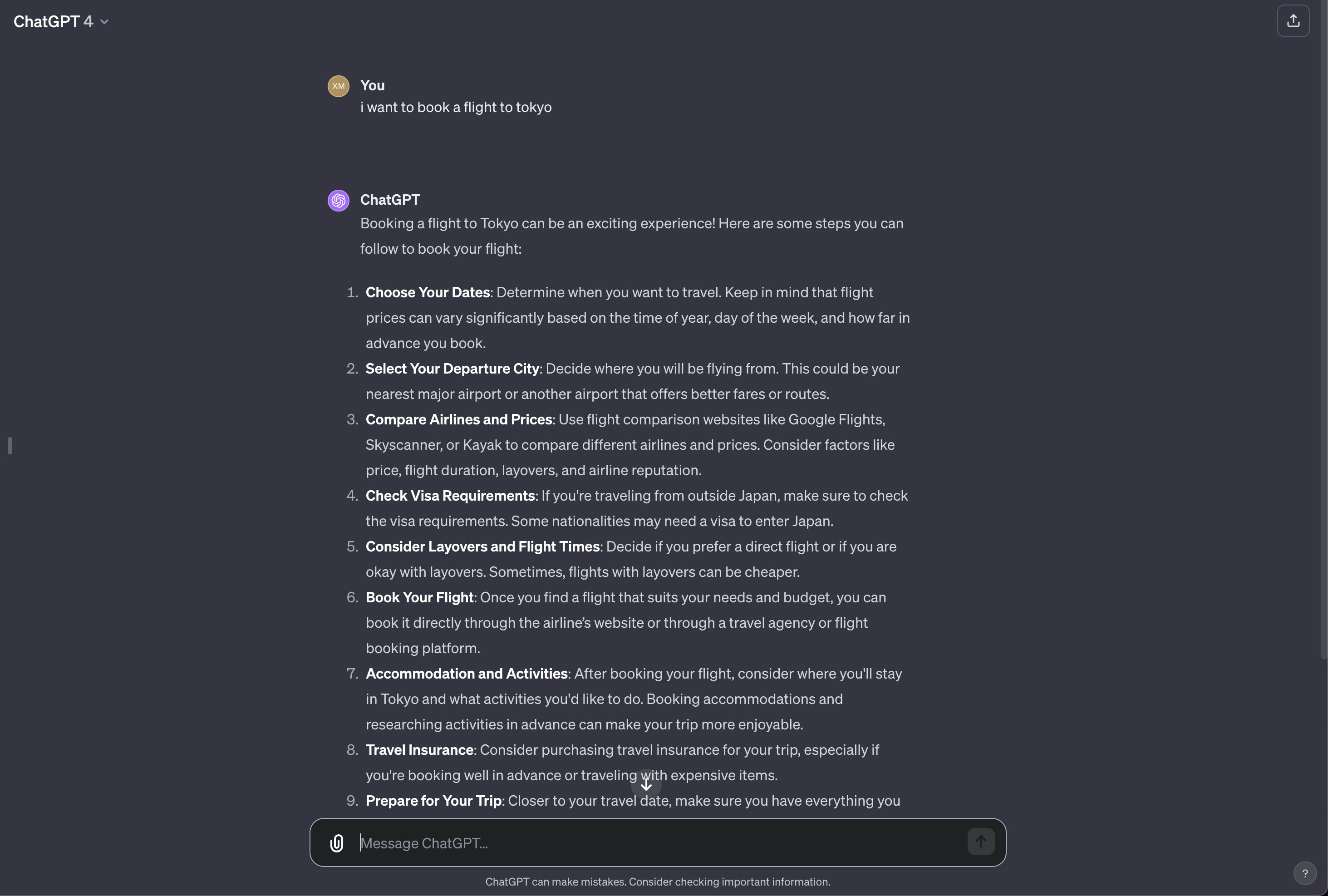}
    \caption{GPT-4 results for the query``I want to book a flight to Tokyo''}
    \label{fig:gpt4_flight}
\end{figure}

\begin{figure}[h]
    \centering
      \includegraphics[width=\linewidth]{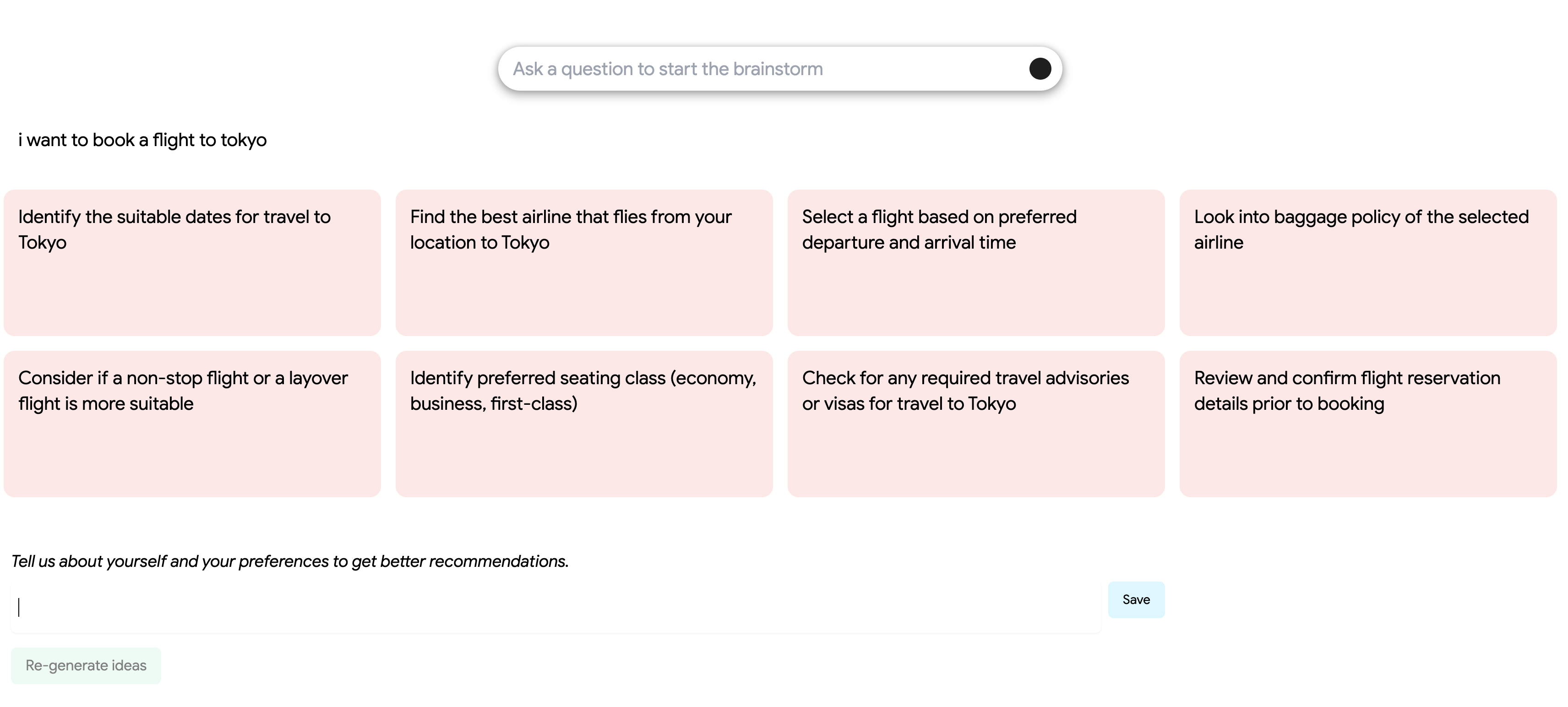}
    \caption{\name{} results for the query ``I want to book a flight to Tokyo''. Users can further interact with each of the sub-tasks.}
    \label{fig:explorellm_flight}
\end{figure}

\begin{figure}[h]
    \centering
      \includegraphics[width=\linewidth]{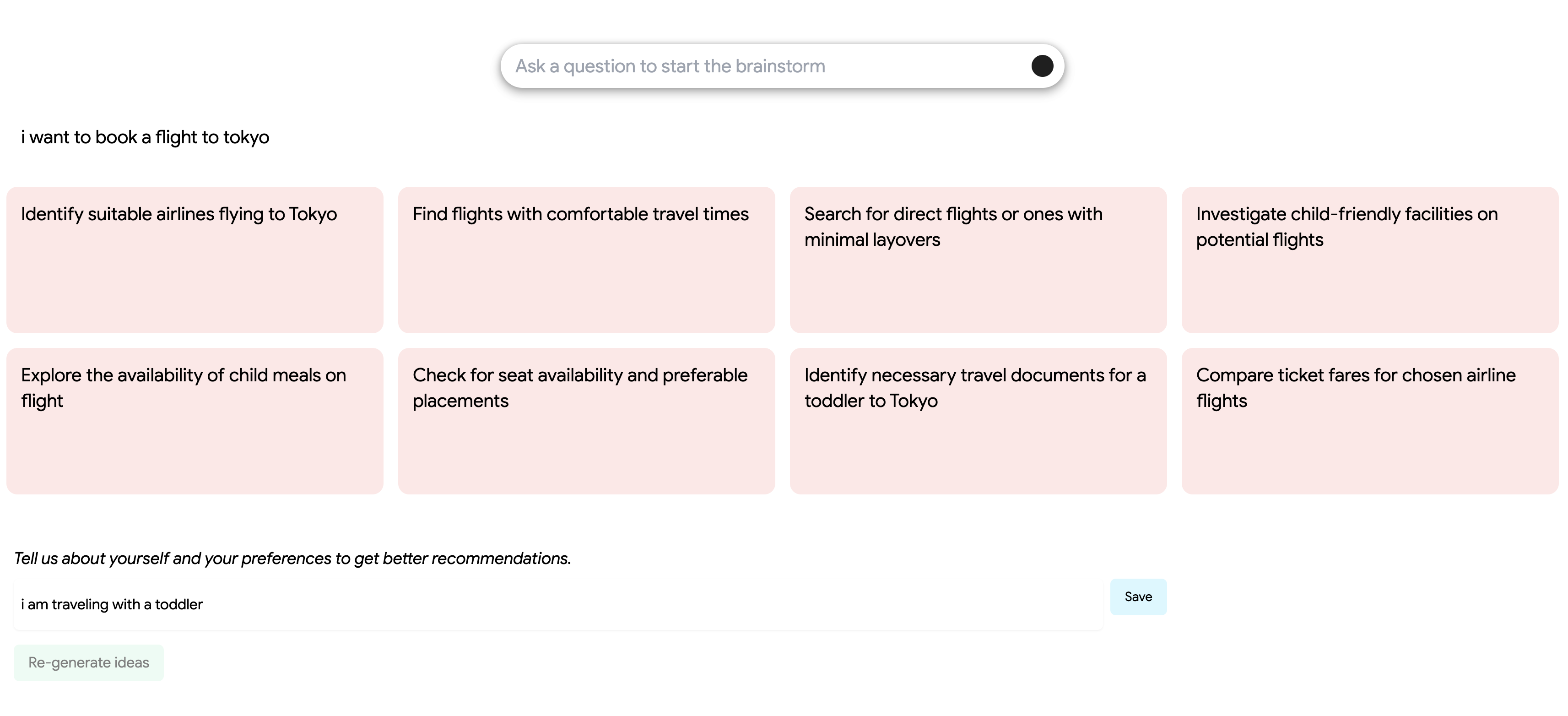}
    \caption{\name{} regenerates and refreshes the sub-tasks for the query ``I want to book a flight to Tokyo'' when the user specifies ``I am traveling with a toddler''.}
    \label{fig:explorellm_flight_personalized}
\end{figure}

\begin{figure}[h]
    \centering
      \includegraphics[width=\linewidth]{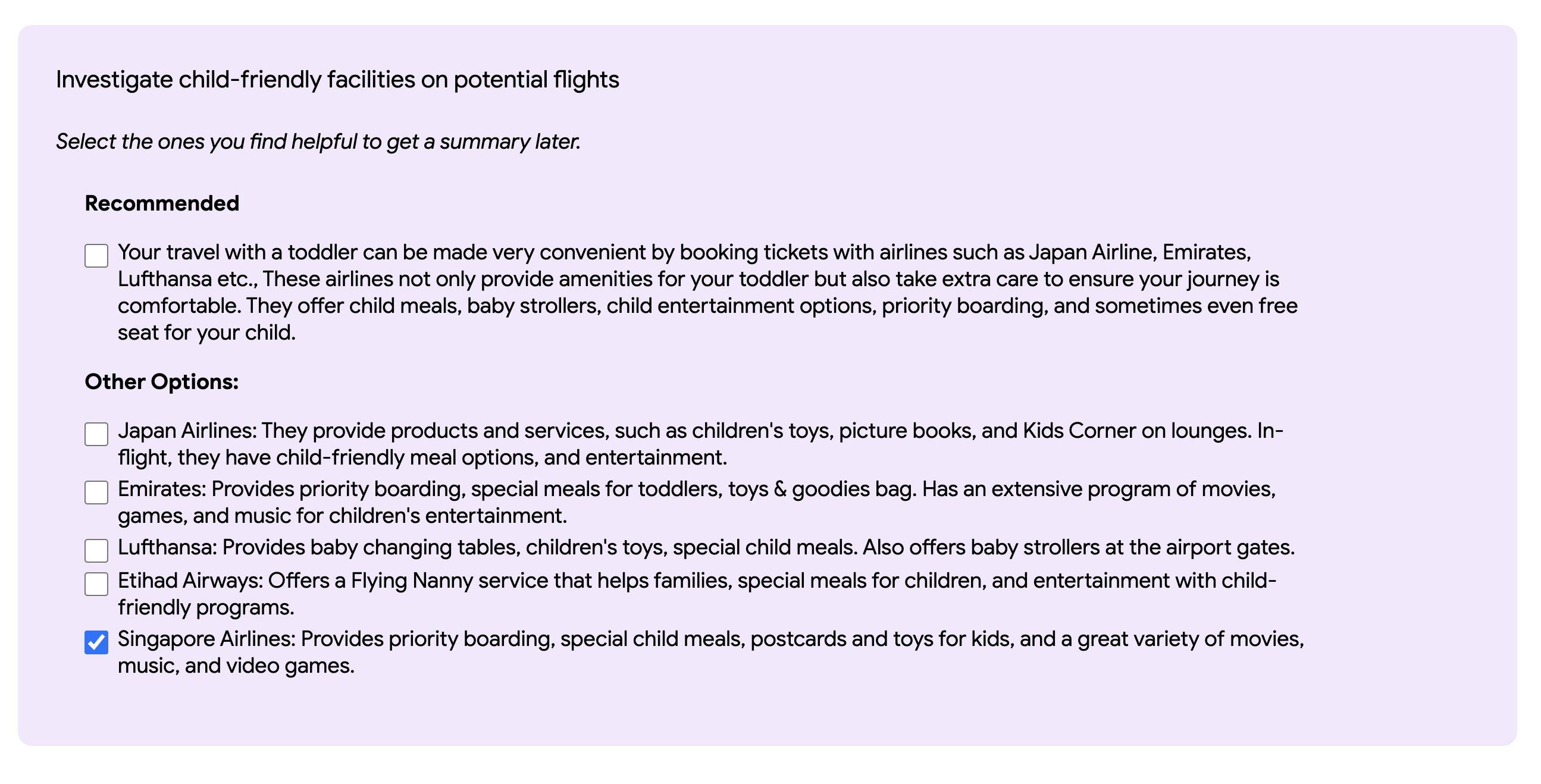}
    \caption{When the user interacts with a subtask such as ``investigate child-friendly facilities on potential flights'', \name{} generates a list of options (may contain hallucination).}
    \label{fig:explorellm_flight_toddler}
\end{figure}

We provide some examples of interaction results with Google Bard (Figure ~\ref{fig:bard_flight}), OpenAI ChatGPT (Figure ~\ref{fig:gpt4_flight}) and \name{} (Figure ~\ref{fig:explorellm_flight},~\ref{fig:explorellm_flight_personalized} and ~\ref{fig:explorellm_flight_toddler}) for reference.

\end{document}